\documentclass[useAMS,usenatbib]{mn2e} 

\usepackage{times,aas_macros,multirow,natbib}
\usepackage{epsfig}
\usepackage{graphicx}

\title[Star-formation around Cl\,0016+16]{A
\textbf{\emph{GALEX}}/\textbf{\emph{Spitzer}} survey of the Cl\,0016+16
supercluster at z=0.55: acceleration of the onset of star-formation in
satellite groups} \author[J.\ E.\ Geach et al.]{J.\ E.
Geach$^{1,2}$\thanks{E-mail: {\tt jimgeach@physics.mcgill.ca}}, R.\ S. Ellis$^{3}$,
Ian Smail$^{2}$, T.\ D.\ Rawle$^{4}$ \& S.\ M.\ Moran$^{5}$\\ 
$^{1}$Department of Physics, McGill University, Ernest Rutherford Building, 3600 Rue University, Montr\'eal, Qu\'ebec, Canada, H3A 2T8\\
$^{2}$Institute for
Computational Cosmology, Durham University, South Road, Durham, DH1\ 3LE,
UK.\\ $^{3}$Department of Astronomy, California Institute of Technology,
Pasadena, CA 91125, USA.\\ $^{4}$Steward Observatory, University of Arizona,
933 North Cherry Avenue, Tucson, AZ 85721, USA.\\ $^{5}$Department of Physics
and Astronomy, Johns Hopkins University, Baltimore, MD 21218, USA.}

\begin{document}

\date{}

\pagerange{ \pageref{firstpage}-- \pageref{lastpage}\ Accepted 29 Nov}
 
\pubyear{2010}

\maketitle

\label{firstpage} 

\begin{abstract} We present the results of a panoramic (15\,Mpc-scale) survey
of the Cl\,0016+16 ($z=0.55$) supercluster using {\it Spitzer Space Telescope}
MIPS 24$\mu$m and {\it Galaxy Evolution Explorer} near-UV (2500\AA; NUV)
imaging. The supercluster regions probed are characterised by several dense
nodes connected by a pronounced intermediate-density filamentary structure.
We have studied the mid-IR and NUV properties of potential cluster members
within a $\Delta z=0.1$ photometric redshift slice, compared to an identical
blank field selection. We have two main findings: (a) the star-formation rates
of individual star-forming galaxies throughout the cluster are not
significantly different to identically selected field galaxies, and (b) the
cluster harbours pockets of `accelerated' activity where galaxies have an
enhanced probability of undergoing star formation. This observation could be
explained in a simple model of `pre-processing' of galaxies during cluster
infall: galaxies in satellite groups have an increased chance of having
star-formation triggered via gravitational tidal interactions compared to
their counterparts in the field, but there is no environmental mechanism
boosting the individual star-formation rates of galaxies. We estimate a
lower-limit for the total star-formation rate of galaxies in the supercluster
as $\sim$850\,$M_\odot$\,yr$^{-1}$ (field corrected). If this rate is
maintained over the typical infall time of a few Gyr, then the infall
population could contribute $\sim$$1$--$2\times10^{12}M_\odot$ of stellar mass
to the structure. \end{abstract} 

\section{Introduction}

One of the characteristics of the formation of structure in the Universe is
the build up of rich clusters via accretion of galaxies, or groups of
galaxies, often arriving in the densest regions through filaments. Since it
has long been known that galaxies falling into rich clusters undergo
significant evolution compared to those remaining in the average density
`field' (Butcher \& Oemler 1974; Dressler\ 1980), what is the role of this
filamentary accretion in influencing the star formation histories of infalling
galaxies? Most studies of cluster environments have concentrated on the
densest environments within them: the cores. However, the intermediate
environments connecting the cores to the surrounding field (or rather, to the
cosmic web) are just as important to study -- especially at $z\sim0.5$ where
the astrophysical processes responsible for shaping the galaxy populations of
the cores of local clusters are operating (Dressler et al.\ 1997).

With wide-field surveys and sensitive multi-wavelength facilities, it is now
becoming more practical to survey these large scale peripheral `interface'
regions. In this work we focus on a survey of the filamentary large scale
structure surrounding the rich cluster Cl\,0016+16 at $z=0.55$. We have used
the Mid-Infrared Photometer (MIPS;\ 24$\mu$m) on board the {\it Spitzer Space
Telescope} and the near-UV (NUV;\ 2500\AA) capabilities of the {\it Galaxy
Evolution Explorer (GALEX)} space telescope to map an extended region around
the cluster, which is one of the best studied environments at intermediate
redshift (e.g.\ Butcher \& Oemler\ 1984; Ellis et al.\ 1997; Dressler et al.\
1999; Brown et al.\ 2000; Worrall \& Birkinshaw\ 2003; Zemcov et al.\ 2003;
Dahlen et al.\ 2004). In particular, Kodama et al.\ (2005) used deep Subaru
SuprimeCam {\it BVRIz} panoramic imaging of the cluster to derive photometric
redshifts over a wide region around the cluster core, showing that two nearby
X-ray selected clusters RX\ J0018.3+1618 and RX\ J0018.8+1602 are connected to
Cl\,0016+16 by a filamentary structure extending some $\sim$20\,Mpc away from
the dense cluster core (Koo 1981; Hughes, Birkinshaw \& Huchra\ 1995; Connolly
et al.\ 1996; Hughes \& Birkinshaw\ 1998). More extensive spectroscopic
follow-up by Tanaka et al.\ (2007) confirms the physical identity of the
filamentary structures.

As in our previous work (Geach et al.\ 2006), our aim is to reveal dusty
star-formation within the cluster, efficiently traced by 24$\mu$m emission
(probing rest-frame 15$\mu$m emission at $z\sim0.5$, a well calibrated
indicator at $z=0$; see also studies by Duc et al.\ 2000, 2004; Fadda et al.\
2000, 2008; Metcalfe et al.\ 2003; Biviano et al.\ 2004; Coia et al.\ 2004;
Geach et al.\ 2006, 2009; Marcillac et al.\ 2007; Bai et al.\ 2007; Oemler et
al.\ 2008; Koyama et al.\ 2008; Dressler et al.\ 2009). In addition to the
{\it obscured} activity, we also exploit the NUV imaging to trace un-obscured
star-formation. The NUV observations allow us to trace star-formation rates
(SFRs) in less obscured (but more numerous) galaxies in the cluster. Tracking
both obscured and un-obscured star-formation is important in building a
complete picture of the environmental influences on galaxy evolution.

This work differs from previous cluster studies at this redshift, because the
physical scale probed by our survey around Cl\,0016+16 allows us to study new
regimes in terms of the dynamic range of local environment. Cl\,0016+16
exhibits the full complement of sub-environments including dense cores,
moderate density filaments and a smooth, intermediate density region blending
in with the surrounding field. Our aim is to perform an unbiased census of the
star-forming populations within these sub-environments to search for
signatures of environmental trends, potentially providing clues on the
mechanisms driving galaxy evolution in the most biased regions at $z\sim0.5$.
In \S2 we describe the {\it Spitzer} and {\it GALEX} observations, and our
main results are presented in \S3. We interpret and discuss the results in
\S4. Throughout this work we assume $h=0.7$ ($H_0 =
100h$\,km\,s$^{-1}$\,Mpc$^{-1}$) and $(\Omega_{\rm
m},\Omega_\Lambda)=(0.3,0.7)$. In this cosmology, the projected scale is
$\sim$300\,kpc/$'$.

\section{Ultraviolet \& mid-infrared observations}

The core region of Cl\,0016+16 was mapped as part of the GTO program 83, and
we expanded this coverage to extend $\sim$$30'\times60'$ in the Cycle 5 GO
program 30263. The AORs were designed to maximise coverage of the extended
filamentary structure identified in the wide-field photometric survey of
Tanaka et al.\ (2007), and match the depth of our existing
MIPS observations of $z\sim0.5$ clusters (Geach et al.\ 2006). Data was
reduced from the Basic Calibrated Data stage using the latest version of the
{\sc mopex} software, and follow the same procedure for post-processing and
object detection as described in Geach et al.\ (2006). We use the coverage and
standard deviation maps generated by {\sc mopex} in the reduction process to
estimate the average depth of the 24$\mu$m map, which has a typical r.m.s. of
$\sim$0.04\ mJy.

We obtained {\it GALEX} imaging of Cl\,0016+16 in September 2007. {\it GALEX}
simultaneously observes co-aligned fields in two ultraviolet bands: FUV
($\lambda_{\rm eff} = 1516$\AA) and NUV ($\lambda_{\rm eff} = 2267$\AA). The
{\it GALEX} field of view is circular, with a nominal diameter of 75$'$,
although the outermost 3$'$ suffers from poorer image quality. The total
exposure time of the observations was ~60\,ksec, giving a limiting sensitivity
in the NUV of 25.6 mag (5$\sigma$). The PSF for both {\it GALEX} bands is
$\sim$5$''$ FWHM, which is well matched to the $\sim$6$''$ MIPS beam.
Photometry was extracted from the standard {\it GALEX} pipeline intensity
maps, measured in 8$''$ diameter apertures. The sky background was estimated
from an annulus with inner radius 5$''$ and outer radius 9$''$.

To isolate galaxies at the redshift of the cluster, we have used the optical
(Subaru) {\it BVRIz} photometric redshift information available over the full
region (Kodama et al.\ 2005). In the analysis of potential cluster members, we
only consider galaxies with $I\leq24$\,mag and with photometric redshifts
selected at $0.5\leq z_{\rm phot}\leq 0.6$ (Tanaka et al.\ 2007). From this
base catalogue, we match $\geq$3$\sigma$ significance 24$\mu$m ($>$0.15\,mJy)
and NUV emitters with a simple geometric $\leq$2$''$ offset criterion. This
results in 58 24$\mu$m detected potential cluster members and 168 NUV detected
potential cluster members respectively.

For the evaluation of environmental trends it is helpful to have a large,
independent, field sample with similar data-sets. For this we turn to the
Cosmological Evolution Survey (COSMOS,\ Scoville et al.\ 2004) -- a 2\,square
degree `blank-field' multi-wavelength survey including Subaru SuprimeCam
optical, {\it SST}/MIPS and {\it GALEX}/NUV imaging. This allows us to perform
an identical selection in an average-biased region, enabling us to normalise
the quantities we evaluate for the cluster. The completeness limits of our
{\it GALEX} and {\it SST} imaging can be well matched to the COSMOS data, and
the relative completeness functions in each band will have a negligible impact
on our result (COSMOS 24$\mu$m 5$\sigma$ limit is 67$\mu$Jy, and the NUV
limiting magnitude is $\sim$26\,AB mag, Sanders et al.\ 2007; Zamojski et al.
2007; P. Capak, private communication). The COSMOS optical imaging was taken
in identical Subaru bands to our survey. By applying identical photometric
selections, we therefore have a very large, robust field control sample; the
Poisson error in the total number of photometrically selected galaxies in the
COSMOS region is $\surd N/N=0.009$

\section{Results}

\subsection{An un-biased view of star-formation in the cluster}

The 24$\mu$m and NUV (probing rest-frame 15$\mu$m and 1613\AA\ at $z=0.55$)
provides a unique new view of star-formation in the cluster, unbiased towards
obscuration. In Figure\ 1 we present the $(V-I)$ vs. $I$ colour magnitude
diagram for potential cluster members (approximately $(U-V)$ vs. $V$
rest-frame). The most interesting feature of this diagram is the clear
bi-modality of the differently selected star-forming populations. The
NUV-selected members, which are expected to trace mainly the un-obscured (or
highly obscured but luminous) star-forming population forms a distinct blue
sequence that extends to $I\sim 23$\,mag. The obscured star-forming systems
selected in the mid-IR have colours best described as `green valley' --
intermediate between the blue cloud and red-sequence, with an average $V-I$
colour, 0.5\,mags redder than the NUV selected members. Note that the
reddening vector moves galaxies roughly perpendicular to the locus of NUV
selected galaxies, and correspondingly there are several 24$\mu$m selected
members scattered into the red-sequence.

This reddening results in a complication in the analysis; while some of these
galaxies may genuinely be very dusty star-formers, evolved passive galaxies on
the red sequence could also emit in the infrared from a buried active galactic
nucleus. These should be excluded from the analysis. Furthermore, the bulk of
galaxies on the red sequence are old, and represent a potentially different
star-formation history to the population we are interested in (the presently
accreting infall population). The field population does not exhibit this
sequence, and so for accurate comparison, we should apply a photometric cut
that excludes them. This is shown as a dashed line on Figure\ 1, of the form
$(V-I) = -0.08 I + 3.5$\,mag.

There are 23 members with detections at both 24$\mu$m and the NUV. Under the
very simple assumption that the UV and mid-infrared emission is generated from
the same star-forming regions in the discs, then these systems provide an
opportunity to assess the total amount of extinction. We apply the calibration
of Geach et al.\ (2006) for 24$\mu$m detected galaxies at $z=0.55$, where the
monochromatic 24$\mu$m luminosity is used to estimate the total
(8--1000$\mu$m) infrared luminosity under the assumption that the SEDs follow
the template models of Dale \& Helou (2002). In this case, ${\rm SFR_{24\mu
m}(M_\odot/yr)} \sim 100f_{\rm 24\mu m}{\rm (mJy)}$. For the NUV fluxes, we
apply the calibration of Salim et al.\ (2007): ${\rm SFR_{FUV}(M_\odot/yr)} =
1.08\times10^{-28}L_{\rm FUV}{\rm (erg\ s^{-1}\ Hz^{-1})}$ (where our NUV
observations are probing the FUV rest-frame, and no {\it k}-correction has
been applied).

For the 23 24$\mu$m and NUV joint detections, we find $\left< {\rm SFR_{24\mu
m} / SFR_{\rm NUV}}\right>\sim17$, but with a large spread: the NUV derived
SFRs underestimate the (presumably true) SFR by a range $\sim$5--75$\times$.
If we naively assume that the disparity can simply be balanced by applying an
extinction correction to the NUV fluxes, then the required reddening to
achieve the average extinction at 1500\AA\ (rest) is 3.1\,magnitudes. Assuming
a $R_V = 4.05$ (Calzetti-type) curve, this corresponds to significant
rest-frame reddening of $A_V = 1.2$\,mag. Note that we have only calculated
the extinction for the NUV and 24$\mu$m co-detections -- we know {\it a
priori} that a significant proportion of ultraviolet photons are able to
escape the galaxy. In reality, as shown in Figure\ 1, while there is some
overlap between the NUV and mid-infrared detected galaxies, there is also
clear bi-modality, implying a distribution of individual extinctions that
extends to more extreme obscuration regime. Similarly, there is a large
population of star-forming galaxies that are not bright at 24$\mu$m. Figure\ 1
shows that the majority of these are likely to be lower mass systems that
trace the majority of the blue-cloud (some of these are even detected in the
FUV band at 1500\AA; evidence of the escape of continuum emission from near
the Lyman limit in the galaxies' rest-frames).

We can perform a basic census of the total star-formation associated with the
cluster. The median SFR of 24$\mu$m and NUV emitters is
25\,$M_\odot$\,yr$^{-1}$ and 1.5\,$M_\odot$\,yr$^{-1}$ respectively. Summing
over all photometrically selected members, we estimate that the total
instantaneous SFR in the supercluster is $\sim$2400\,$M_\odot$\,yr$^{-1}$.
Corrected for the contribution from the field, which we estimate from the
average SFR surface density in the COSMOS field for an identically selected
sample, we find a total SFR of $\sim$850\,$M_\odot$\,yr$^{-1}$, which is
overwhelmingly dominated by the infrared output (the NUV component has not
been corrected for intrinsic extinction, and therefore this value can be
considered a lower limit). The total SFR in the structure is in the range seen
in infrared surveys of other massive clusters at similar redshift, which are
characterised by a mass normalised rate of
$\sim$10--100\,$M_\odot$\,yr$^{-1}$\,$(10^{14}M_\odot)^{-1}$ (Geach et al.\
2006)\footnote{The core mass of Cl\,0016+16 is in excess of $10^{15}M_\odot$,
but here we have surveyed a large volume around the extended structure which
encompasses a larger mass -- therefore direct comparisons between clusters
should be made with caution.}. If the activity continues at a constant rate
over the infall time-scale (few Gyr), then this population could contribute
$\sim$$1$--$2\times10^{12}M_\odot$ stellar mass to the descendant of this
environment. Are individual SFRs of galaxies in the cluster significantly
different to the field? To test this, we perform a Kolmogoroff-Smirnov (K--S)
test of the NUV and 24$\mu$m flux distributions in equivalent photo-$z$ slices
in the Cl\,0016+16 field and the COSMOS field. We find no significant
difference in either the NUV or 24$\mu$m selected galaxies, with K--S
probabilities of 0.999 and 0.980 respectively.

\begin{figure} \includegraphics[width=0.47 \textwidth]{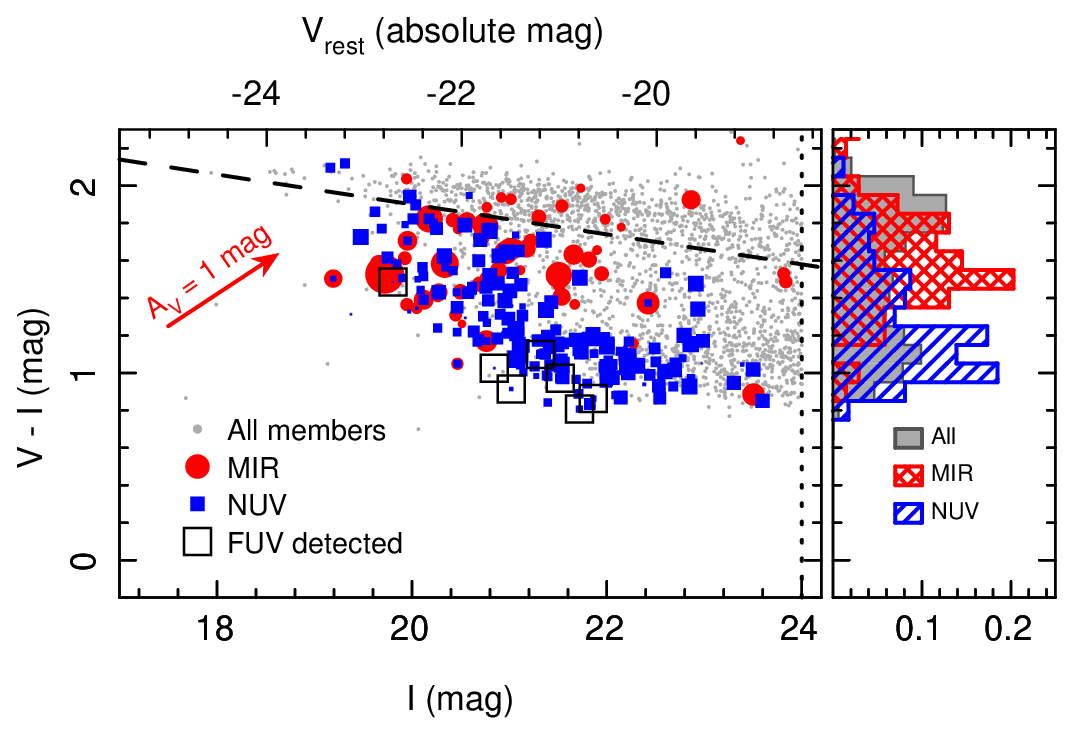} \caption{$(V-I)$
colour magnitude diagram of galaxies with $0.5\leq z_{\rm phot}\leq0.6$ in the
Cl\,0016+16. The full population is shown (grey points), and 24$\mu$m and NUV
detected members overlaid, scaled by flux density. The histogram provides a
clearer indication of the colour distribution, with NUV emitters dominating a
sequence of blue galaxies, with the 24$\mu$m emitters generally $\sim$0.5 mags
redder in $(V-I)$. Note that the reddening vector (where $A_V$ is the
extinction in the rest-frame $V$) moves galaxies roughly perpendicular to the
blue sequence -- this can account for some mid-infrared detections in the
red-sequence of Cl\,0016 (although active galactic nuclei in evolved massive
hosts could also contribute, and these should also be excluded from the
analysis). The dashed line describes a simple photometric cut to eliminate the
red-sequence from our analysis of the infall population (\S3.1).} \end{figure}

\begin{figure*} \includegraphics[width=0.4 \textwidth]{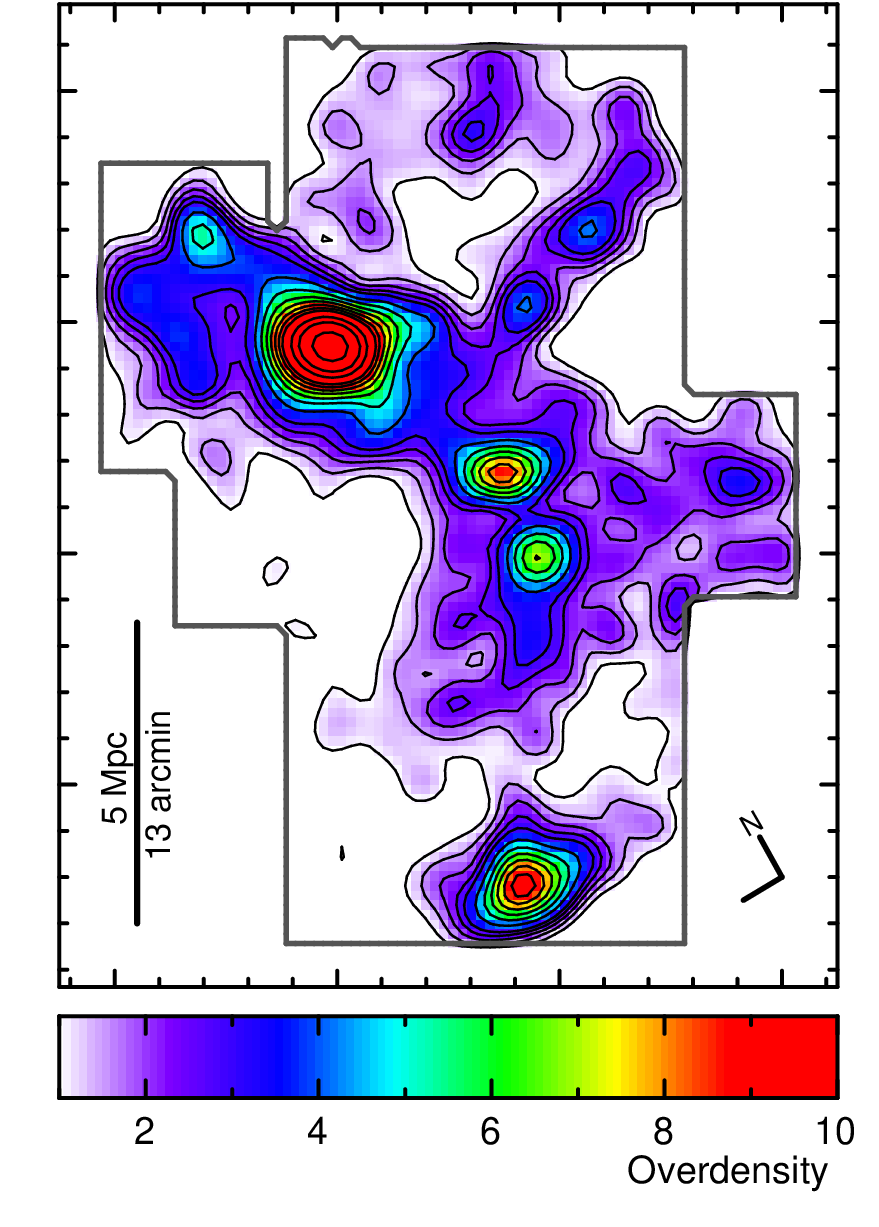}
\includegraphics[width=0.4 \textwidth]{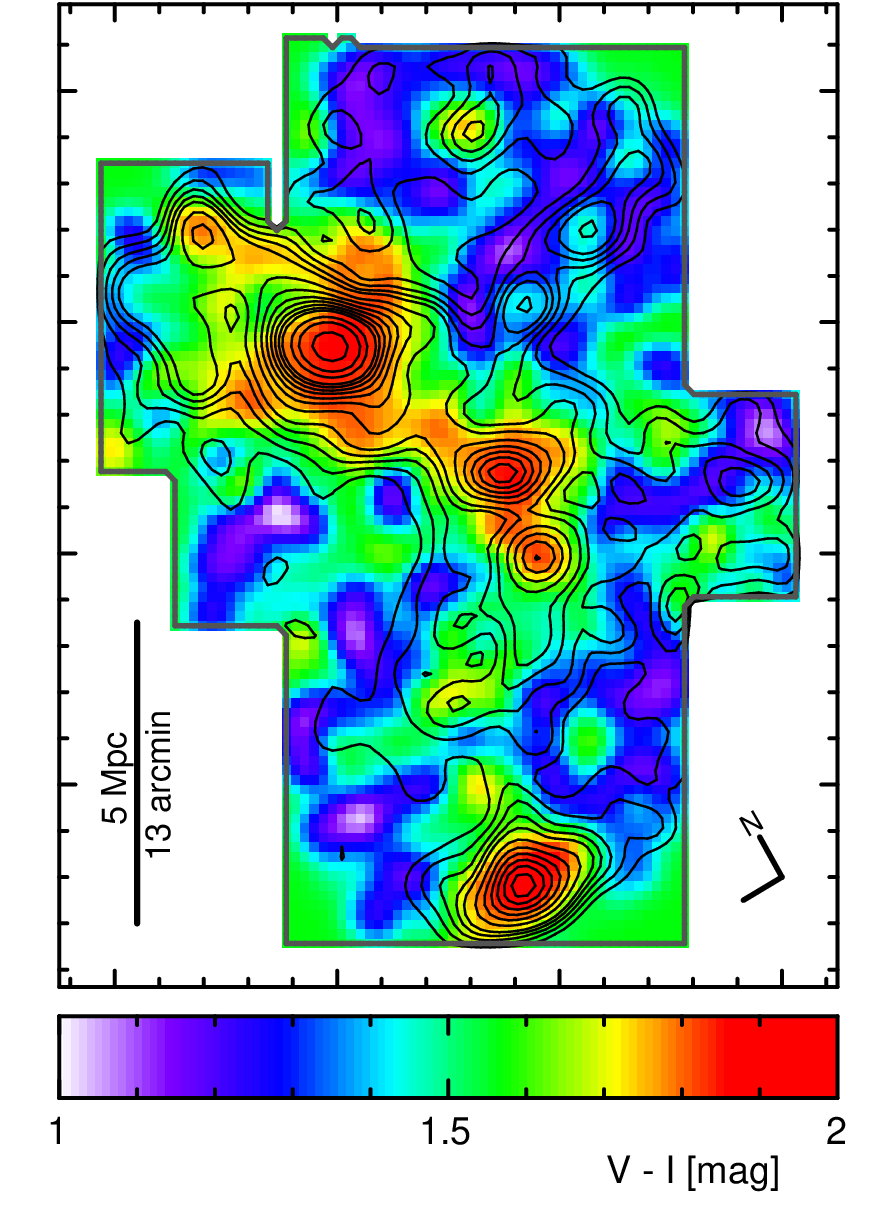} \includegraphics[width=0.4
\textwidth]{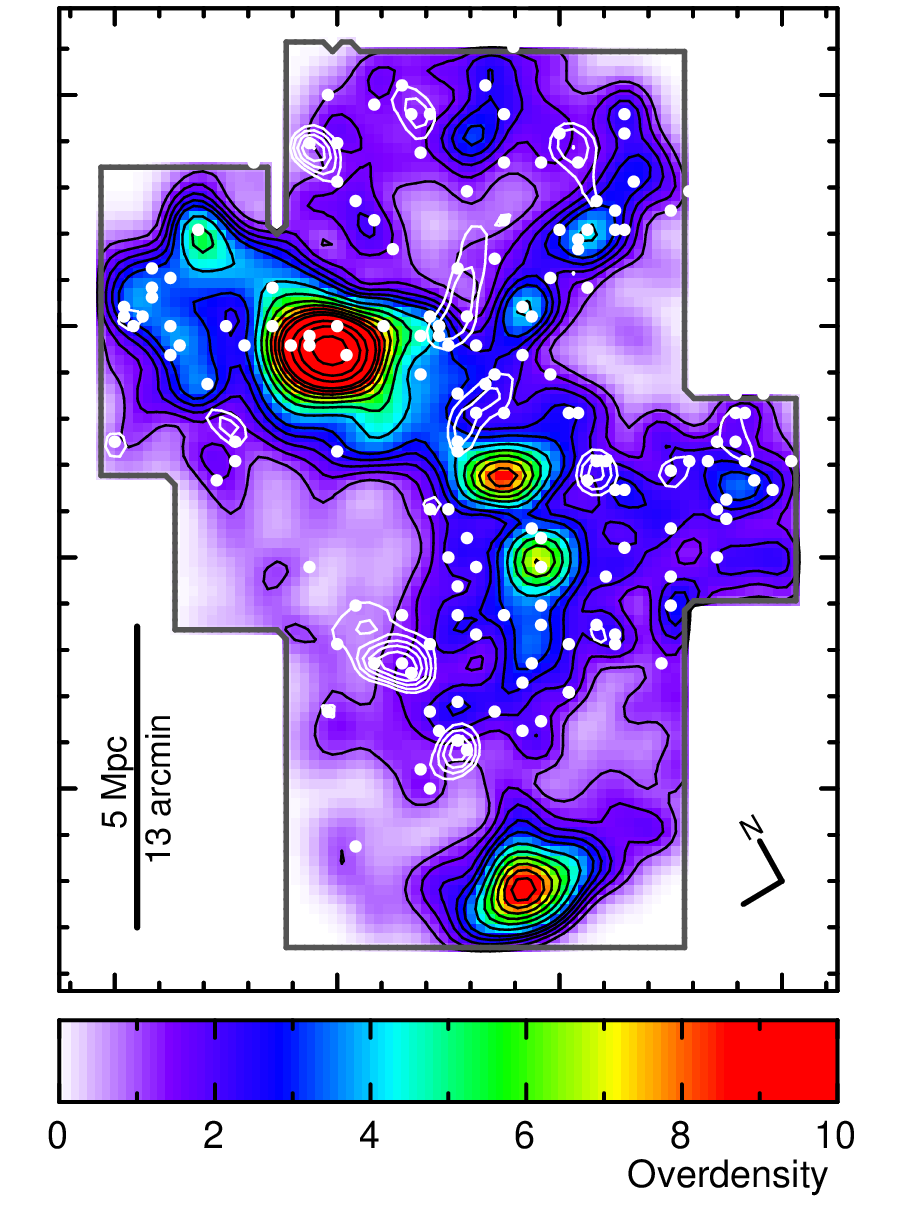} \includegraphics[width=0.4 \textwidth]{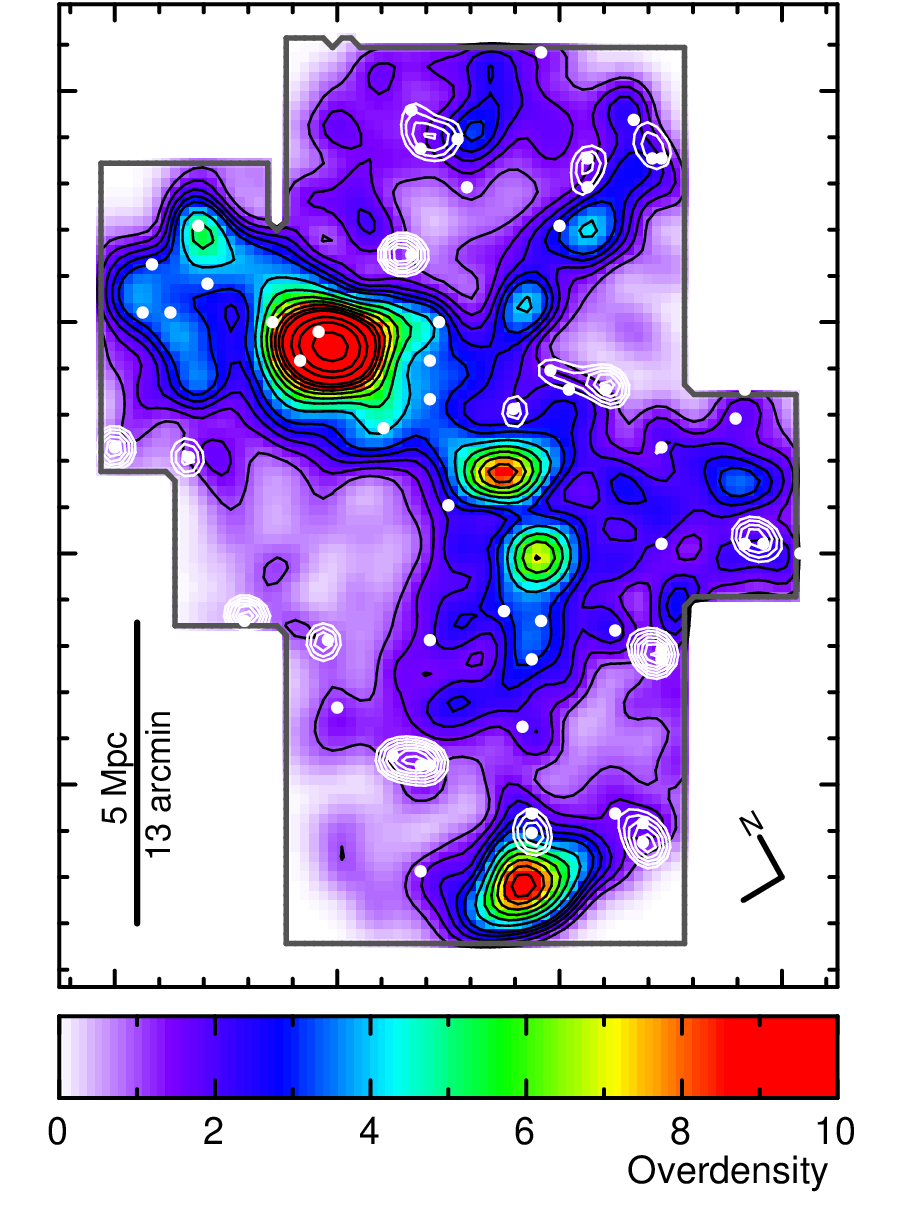}
\caption{Maps of the Cl\,0016+16 filamentary structure; the main cluster core
is at 00$^{\rm h}$18$^{\rm m}$30.4$^{\rm s}$,
+16$^{\circ}$26$^{'}$17.3$^{''}$. (top left) the surface overdensity (relative
to the COSMOS field) of galaxies selected at $0.5\leq z_{\rm phot}\leq0.6$.
Contours start at unity and are spaced at half-integer intervals. (top right)
Average $(V - I)$ colour of galaxies: the dense cores are characterised by red
galaxies, but the lower-density outskirts (including the pronounced filament)
have blue/`green-valley' colours. (bottom) distribution of (left) NUV and
(right) 24$\mu$m emitters shown as points. We have defined a fractional excess
-- $\psi$ (see \S\S3.2) -- which highlights regions where the probability of
any one galaxy in the local vicinity is significantly enhanced compared to the
field (white contours). Although star-forming galaxies can be seen distributed
throughout the structure, these regions could be interpreted as pockets of
accelerated star-formation where activity has been triggered within
group-scale satellite structures being accreted onto the cluster.}
\end{figure*}

\subsection{The distribution of star-formation in the cluster}

Since we are mainly interested in the distribution (and field comparison) of
the star-forming galaxies relative to the overall large-scale structure, we
have created spatially smoothed maps of the cluster\footnote{In this mapping
analysis, we exclude the quasar within the cluster ($z=0.553$) at ($\alpha$,
$\delta$) = (00$^{\rm h}$18$^{\rm m}$31.9$^{\rm s}$, 16$^{\rm
\circ}$29$'$26$''$) (Margon et al.\ 1983). The QSO is co-detected in the
far-UV, near-UV and 24$\mu$m bands; the total far-UV (1500\AA), near-UV
(2500\AA) and 24$\mu$m flux density of the QSO are 59$\mu$Jy, 91$\mu$Jy and
2.5mJy respectively.}. Taking the MIPS image as a starting point, we first
created a binary mask image based on the 24$\mu$m coverage. We adopt a grid
scale of 25$''$/pixel, and at every grid cell we estimate the surface density
of all galaxies within 1$'$ ($\sim$300\,kpc). We map the local surface density
of all photometric redshift selected galaxies across the field by counting the
number of galaxies within 300\,kpc and normalising to the density of
identically selected galaxies in the COSMOS field. Figure\ 2 shows the
resultant `overdensity' map of the super cluster, highlighting several dense
nodes, surrounded by an intermediate-density environment with average surface
densities 2--3$\times$ that of the field. For comparison, we also show the
local average $\left< V-I\right>$ colour of galaxies in corresponding
environments; clearly showing the pronounced reddening of galaxies towards
regions of high density.

We introduce a simple measure to map how star-formation is distributed across
the structure, again evaluated on 300\,kpc scales: \begin{equation} \psi =
\frac{\rm Local\ fraction\ of\ star\ forming\ cluster\ members}{\rm Average\
fraction\ of\ star\ forming\ field\ galaxies} \end{equation} Where local
fraction of star-forming cluster members can be taken as the ratio of the
number of galaxies above a limiting NUV or 24$\mu$m flux to the total number
of galaxies selected in the photometric redshift slice. By normalising to the
equivalent fraction in the field, a map of this simple statistic should reveal
where galaxies have an increased {\it probability} of exhibiting on-going
star-formation. This is a measure of excess activity that takes into account
the fact that there are naturally more galaxies in the cluster environment.

Maps of $\psi$ calculated for the NUV and 24$\mu$m emitters are shown as
contours in Figure\ 2. Regions where $\psi>1$ can be considered as
environments where the probability of any one galaxy in the locale is enhanced
over the surrounding regions. Obviously, this statistic is sensitive to
shot-noise, and so the significance of values of this `excess' must be
carefully assessed. We do this by performing a boot-strap like simulation,
re-evaluating $\psi$ after repeatedly randomly shuffling the NUV and 24$\mu$m
fluxes of potential cluster members. The fraction of pixels in the randomised
map with some $\psi>\psi'$ is an indication of the significance of finding
regions with similar excess in the `real' map. We perform this randomisation
100 times to build-up a robust estimate of the distribution of $\psi$ in the
shuffled maps, and define significant values as $P(\psi)\leq0.01$. This
corresponds to values of $\psi=1.2$ and $\psi=2.6$ in the NUV and 24$\mu$m
respectively, and we fix these as the lowest contours in Figure\ 2.

The K--S test of the flux distribution described in \S\S3.1 showed that there
is no significant difference between the SFRs of galaxies in and around the
cluster and the randomly sampled field. This implies that there is {\it no
environmental process boosting the SFRs of individual galaxies}. However, we
have identified regions within the cluster where {\it the probability of a
galaxy undergoing star-formation is enhanced}. Although these pockets of
activity do not account for all of the star-formation currently occurring in
the cluster (Figure\ 2 shows that NUV and 24$\mu$m emitters are detected
throughout the structure), they point to a important stage in the
star-formation histories of galaxies evolving in highly biased environments.
It should not surprise us that the {\it average} SFRs of these galaxies are
similar to the field. Unless the galaxies being accreted onto the cluster are
systematically richer in their molecular gas content, there is nothing {\it
intrinsic} that would boost their SFRs once triggered. A caveat to this is
that it could be possible to drive star-formation at higher rates via more
intense local environmental processes: mergers or violent tidal interactions
for example -- these can perturb and collapse the molecular gas clouds within
discs, perhaps boosting the density (and therefore intensity) of star-forming
regions within them. Without more detailed follow-up study of these groups
(for example high-resolution imaging, or a molecular gas survey for galaxies
at fixed stellar mass in the cluster and the field), it is difficult to
conclusively say what the dominant triggering mechanisms are, but we propose
that mild galaxy-galaxy tidal interactions are likely to play a pivotal role
(Martig \& Bournaud\ 2008). The increased frequency of these tidal events
within small satellite groups as they are accreted into the cluster could be
responsible for the observed enhancement of the accelerated onset of
star-formation in these galaxies.

\section{Discussion \& Conclusions}

The build-up of stellar mass in rich clusters is intimately linked to the
accretion of groups. Semi-analytic models that apply a comprehensive treatment
for the environmental processing of galaxies as they are accreted into massive
halos (Font et al.\ 2008) predict that a large fraction of stellar mass
assembly of massive ($>$$10^{14}M_\odot$) halos comes from the merging of
group-size halos. In an analysis of stellar mass assembly following halo
merger trees underpinning a {\sc galform} semi-analytic simulation, McGee et
al.\ (2009) show that at $z=0.5$, 30--40\% of the total stellar mass in halos
of mass $10^{{14}-{15}}M_\odot$ was in a substructure of mass of at least
$10^{13}M_\odot$ at the time of accretion. Thus, a large fraction of the
stellar mass budget of rich clusters is delivered via the infall of groups.
The panoramic observations of the supercluster around Cl\,0016+16 presented
here offer a snapshot of this process in action. Where star formation is
occurring, individual rates appear to be at the same level as an identically
selected field-sample; there is apparently no process that is systematically
boosting the SFRs. However, we have identified pockets of activity in the
periphery regions which can be interpreted as an enhancement of the {\it
probability} of any one galaxy exhibiting on-going star formation. This could
be a sign of an acceleration of the onset of star-formation in infalling
satellite groups as they are assimilated into the supercluster; a process that
is likely to take several Gyr.

Our results are in good agreement with other similar mid-infrared surveys of
filament structures at higher and lower redshifts. For example, in another
24$\mu$m survey, Fadda et al.\ (2008) show that star formation is enhanced
(relative to the rest of the cluster) in filaments feeding the cluster Abell
1763 ($z=0.23$). It is argued that the inefficiency of the hot intracluster
media of filaments allows galaxies to retain their cold gas reservoirs
(compared to the much harsher conditions in the inner regions), and the
relatively low velocity dispersions could promote galaxy-galaxy tidal
interactions that may trigger star formation during infall. At higher
redshift, $z\sim0.8$, Marcillac et al.\ (2007), Bai et al.\ (2007) and Koyama
et al.\ (2008) use mid-infrared observations to show that the outskirts of
more distant clusters also contain large populations of starburst galaxies,
but that this activity is suppressed, or at least, is not obvious in the
clusters' highest density regions. More recently direct far-infrared and
submillimeter cluster surveys have also revealed the presence of an active
population of galaxies on the peripheral environments of rich clusters (e.g.\
Braglia et al.\ 2010, Rawle et al.\ 2010, Pereira et al.\ 2010), consistent
with this picture.

What physical process is driving this form of enhancement? A simple scenario
is that the star-formation is being triggered during mild tidal interactions
during the assembly of small, bound groups. Given the long time-scale for the
complete accretion of a satellite group into the cluster core, this implies
that a significant fraction of the stellar mass assembly actually occurs
`in-place' within the groups. Indeed, other observational studies have shown
that there is a enhanced fraction of early-type `red' galaxies in groups of
galaxies (Zabludoff \& Mulchaey\ 1998; Tran et al.\ 2001), implying at least
some accelerated evolution of galaxies in even moderate environments compared
to the field. The inefficiency of the intragroup medium for terminating
star-formation via ram pressure stripping, and related processes (e.g.\ Moore
et al.\ 1996, Bekki\ 2009) means that once triggered, there is little to
hinder the activity, and the timescale for cluster infall is long enough for
significant stellar evolution to take place (several\,Gyr, e.g.\ Treu et al.\
2003). Thus, the cessation of star-formation in the cluster outskirts is more
likely to be governed by gas exhaustion -- long before the galaxies reach the
more hostile cluster core. This does not mean that the virialised cluster
environment does not profoundly influence these galaxies' star-formation
histories. Although stellar mass can be built up without hindrance on the
cluster outskirts, once the galaxies reach the hotter, virialised intracluster
medium of the core, further cooling of gas (and therefore further stellar mass
build-up) is prevented.

The next step is to understand the group-scale physics responsible for
triggering star-formation, and perform a more detailed study of the mode of
star-formation in these galaxies. One of the most important goals is to
understand more subtle dependencies on cluster galaxies' star-formation
histories that connect both the galaxies' intrinsic properties (most
importantly their mass) and wider environmental effects. As we have shown
here, such studies require the union of large, representative field samples
and panoramic multi-wavelength surveys of the rare cluster environments.

\section*{acknowledgements}

We thank the referee for a constructive report. The authors wish to thank
Masayuki Tanaka, Taddy Kodama \& Yusei Koyama for providing a merged catalogue
of spectroscopic and photometric redshifts for the extended cluster region,
Subaru SuprimeCam (PISCES) imaging and useful comments, and Peter Capak for
helpful discussions on the COSMOS data. J.E.G. acknowledges the National
Science and Engineering Council (NSERC) of Canada and the U.K. Science and
Technology Facilities Council (STFC). I.R.S. also acknowledges STFC. J.E.G.
thanks the Royal Society for the award of an International Travel Grant, and
the hospitality of the California Institute of Technology during March 2010,
which expedited the completion of this work. This research has been based on
observations made with the {\it Spitzer Space Telescope} and the {\it Galaxy
Evolution Explorer} which are operated by the Jet Propulsion Laboratory,
California Institute of Technology, under a contract with NASA.

\label{lastpage}

\end{document}